\documentclass[preprint,authoryear,12p,twocolumn]{article}

\usepackage{booktabs}
\usepackage{verbatim}
\usepackage{xspace}
\usepackage{subcaption}
\usepackage{todonotes}
\usepackage{natbib}

\graphicspath{{img/}}

\usepackage{siunitx}
\DeclareSIUnit\erg{erg}
\DeclareSIUnit\rad{rad}
\DeclareSIUnit\years{a}
\DeclareSIUnit\hours{hours}
\DeclareSIUnit\%{\percent}
\DeclareSIUnit\unity{none}
\DeclareSIUnit\au{AU}
\DeclareSIUnit\arb{arbitrary units}
\DeclareSIUnit\pixel{px}
\DeclareSIUnit{\promille}{\textperthousand}
\DeclareSIUnit\parsec{pc} 
\DeclareSIUnit\lightyear{ly}
\DeclareSIUnit\LD{LD}

\usepackage[%
breaklinks=true,%
colorlinks=true,%
pdfauthor={Bambach et al.},%
pdftitle={Template for manuscripts in Advances in Space Research}%
]{hyperref}

 \usepackage[nameinlink,noabbrev,capitalize]{cleveref}

\newcommand{\tabitem}{~~\llap{\textbullet}~~}
\newcommand{\steins}{\texorpdfstring{\v{S}teins\xspace}{Steins\xspace}}
\newcommand{\deltav}{\ensuremath{\Delta v}\xspace}
\begin{document}
\twocolumn[{
\begin{flushleft}
{\Large\textbf\newline DISCUS - The Deep Interior Scanning CubeSat mission to a rubble pile near-Earth asteroid}

{Patrick Bambach}\textsuperscript{1*}
{Jakob Deller}\textsuperscript{1}
{Esa Vilenius}\textsuperscript{1}
{Sampsa Pursiainen}\textsuperscript{2}
{Mika Takala}\textsuperscript{2}
{Hans Martin Braun}\textsuperscript{3}
{Harald Lentz}\textsuperscript{3}
{Manfred Wittig}\textsuperscript{4}
\\
\bigskip
\bf{1} Max Planck Institute for Solar System Research, Justus-von-Liebig-Weg 3, 37077 G\"ottingen, Germany
\\
\bf{2} Tampere University of Technology, PO Box 527, FI-33101 Tampere, Finland
\\
\bf{3} RST Radar Systemtechnik AG, Ebenaustrasse 8, 9413 Oberegg, Switzerland
\\
\bf{4} MEW-Aerospace UG, Hameln, Germany
\\
\bigskip
* \url{bambach@mps.mpg.de}
\\
{\textit{Submitted to Advances in Space Research}}



\noindent\rule{\textwidth}{1pt}

\end{flushleft}
\begin{abstract}

We have performed an initial stage conceptual design study  
for the {\em Deep Interior Scanning CubeSat} (DISCUS), a tandem 6U CubeSat carrying a bistatic radar as the main payload. 
DISCUS will  be operated either as an independent mission or accompanying a larger one. It is designed to determine 
the internal macroporosity of a 260--600 m diameter Near Earth Asteroid (NEA) from a few kilometers distance. 
The main goal will be to achieve a global penetration with a low-frequency signal as well as to analyze the scattering strength for various different penetration depths and measurement positions. Moreover, the measurements will be inverted through a computed radar tomography (CRT) approach. The scientific data provided by DISCUS would bring more knowledge of the  internal configuration of rubble pile asteroids and their collisional evolution in the Solar System. It would also advance the design of future asteroid deflection concepts. We aim at a single-unit (\si{1}{U}) radar design equipped with a half-wavelength dipole antenna. The radar will utilize a stepped-frequency modulation technique the baseline of which was developed for ESA's technology projects GINGER and PIRA. 
The radar measurements will be used for CRT and shape reconstruction. 
The CubeSat will also be equipped with an optical camera system and laser altimeter to support navigation and shape reconstruction. We  provide the details of the measurement methods to be applied along with the requirements derived of the known characteristics of rubble pile asteroids. Additionally, an initial design study of the platform and targets accessible within
20 lunar distances is presented.

\end{abstract}

\textit{Keywords:}\\
Deep-space CubeSat,\quad{} near earth asteroids,\quad{} rubble pile asteroid,\quad{}  radar,\quad{}   computed radar tomography 

\noindent\rule{\textwidth}{1pt}
\vspace{2em}
}]
\section{Introduction}

\begin{figure*} 
\centering
  \includegraphics[width=0.45\textwidth]{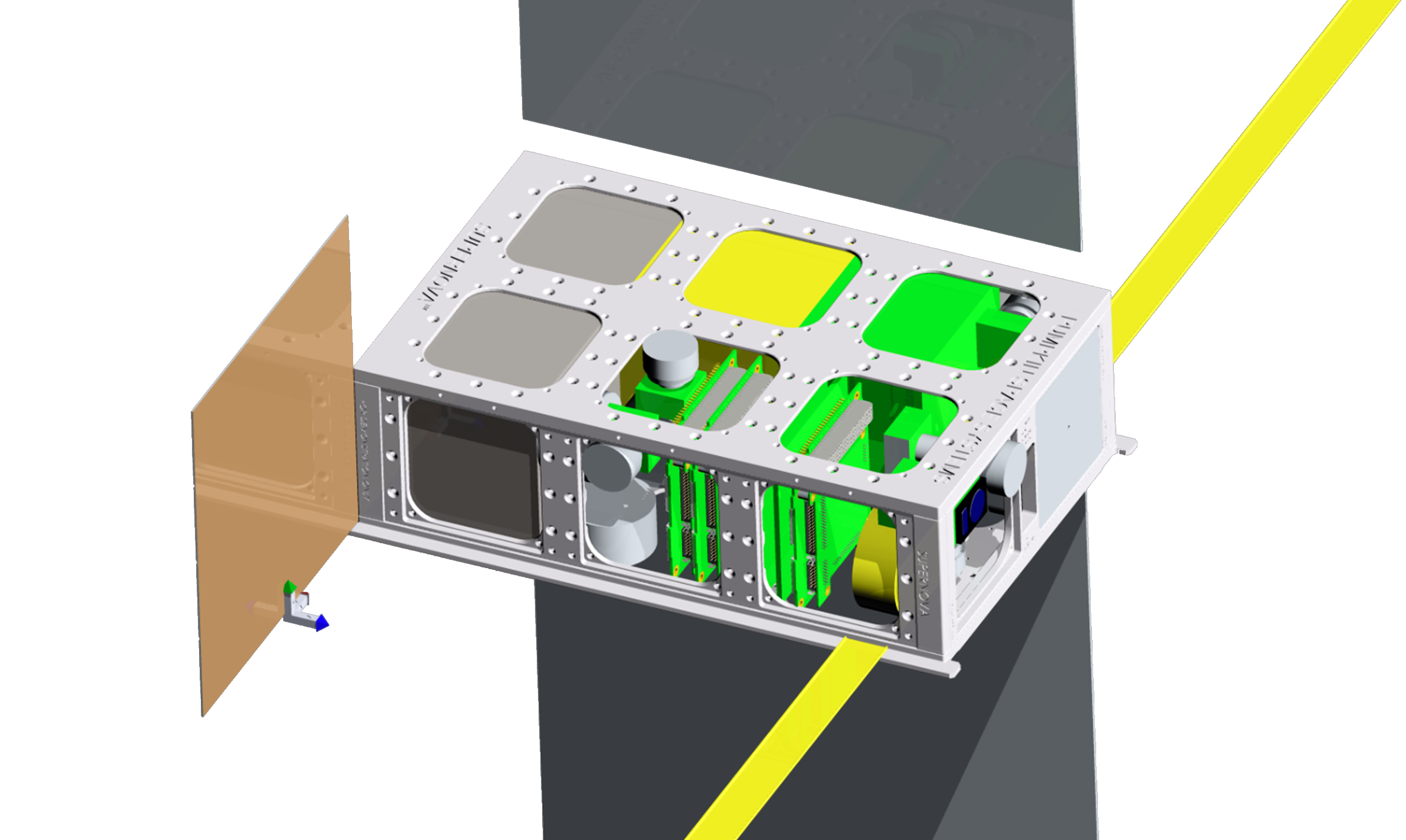} 
  \includegraphics[width=0.5\textwidth]{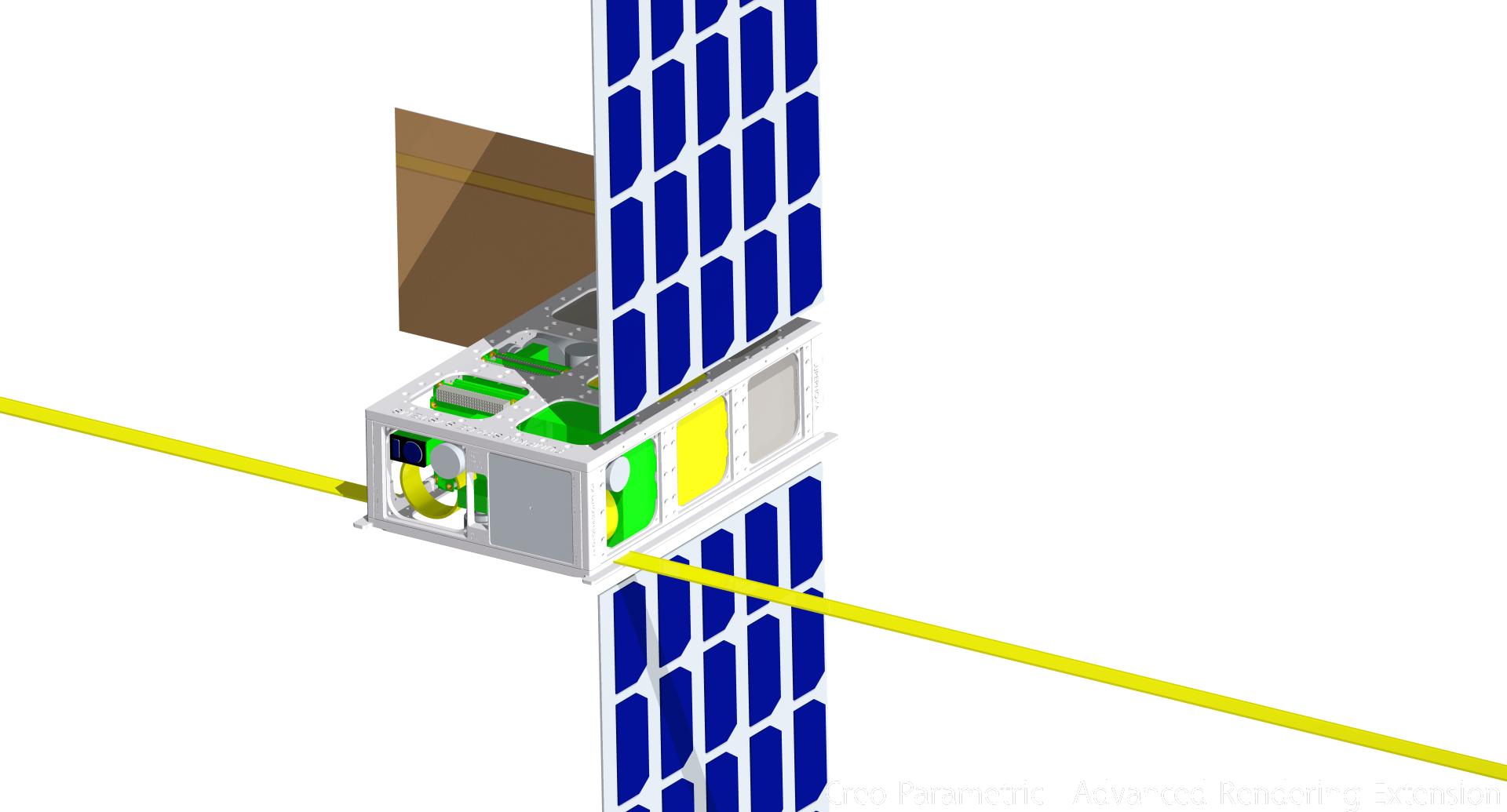}
\caption{Conceptual pictures of the 6U DISCUS design  containing  (a) 1U radar (green), (b) 1U Communication (yellow), (c) 2U BIT-3  propulsion system (gray).  Attached to the frame is a linear  half-wavelength dipole antenna (yellow line). During the launch, the antenna will be rolled up. The other parts are components of the shelf (COTS). Among others a Pumpkin 6U frame with four star trackers, an  altimeter, a camera, three reaction wheels and PCB boards for housekeeping and a cold gas thruster. Two solar panels are oriented in a perpendicular direction with respect to the antenna.  The radar measurement is performed with the antenna pointed towards the sun in order to minimize the noise caused by solar radiation. It also maximizes the power gain of the solar panels and reduces the need for power storage. }\label{fig:discus_6U}
\end{figure*}

The goal of this paper is to advance the mission design for recovering the deep interior structure of an asteroid using small spacecraft. We introduce the initial stage mission concept, the {\em Deep Interior Scanning CubeSat} (DISCUS),  in which the goal is to fly two identical six-unit (6U)  CubeSats (\cref{fig:discus_6U}) as a tandem into the orbit of a small rubble pile asteroid and to resolve its global interior structure via tomographic radar measurements. This is an important scientific objective which has been approached via several mission concepts \citep{safaeinili2002,asphaug2003,herique2016,kofman2007,snodgrass2017}. In particular, we approach the {\em Deep Interior} concept by \cite{asphaug2003} as a state-of-the-art small spacecraft mission \citep{NASASmallSpacecraft} which can be either  independent or accompanying a larger one. DISCUS CubeSats will be equipped with a bistatic penetrating radar and an optical imaging system. The design is derived from airborne Ground Penetrating Radar (GPR) which is today applied in mapping subsurface glacier and soil structures, e.g., ice thickness \citep{rutishauser2016,gundelach2010,eisenburger2008,fu2014}. 

The first realized tomography attempt for small a Solar System body was the COmet Nucleus Sounding Experiment by Radio-wave Transmission (CONSERT) \citep{kofman2007, kofman2015} during the European Space Agency's (ESA) {\em Rosetta} mission in 2014. In CONSERT, a radio-frequency signal was transmitted between the {\em Rosetta} orbiter and its lander {\em Philae}. Rosetta was a major scale mission with its total budget of  1.4~Billion Euro with the cost of the lander being 200 Million. Even thought the missions has been a success the high costs hinders the development of a comparable successor on this scale. 
Due to lower cost and the recent technological advances, small spacecraft would enable the exploration of both asteroids and comets more flexibly and in higher numbers.
For comparison, the recently published M-ARGO concept aims for a budget of 25~Million Euro \citep{M-ARGO}.
The interest towards using CubeSats to complement traditional planetary missions is constantly increasing and, therefore, mission design to support this development is needed. A few years ago, \cite{6U_Interplanetary} published an initial report on potential Interplanetary CubeSats with 2U  science payload and 2U propulsion module. Later on in 2014, NASA announced with NeaScout  the first detailed mission design of a 6U CubeSat to fly with a solar sail to an asteroid \citep{NeaScout}. Recently developed CubeSats capable of operating in deep space  include the Lunar IceCube \citep{clark2016} and Mars Cube One \citep{asmar2014,rahmat2017,hodges2016}. 

In \cite{JPL_Components}, the current status of 6U Interplanetary CubeSat development at the Jet Propulsion Laboratory (JPL) is given. Critical hardware, such as data handling and communication systems have been developed as standard buses at JPL. Namely, the Sphinx data handling system and the Iris transponder. The volume and weight of this hardware show that the promise of 6U interplanetary CubeSats with around 2U housekeeping, 2U propulsion and almost 2U payload could be kept. An ESA/ESTEC study found the limitations of 6U for a NEA mission to be critical, especially regarding on thermal management and \deltav requirements of independent interplanetary missions. Therefore, a 12U based CubeSat concept M-ARGO \citep{M-ARGO} has been investigated and found promising. Besides completely autonomous missions also so-called CubeSat piggy-bag missions can become more common in the future. In the piggy-bag approach, a mothership transports the CubeSats to its target, deploys them and serves as a communication link. The advantage  is that the requirements for navigation, communication, propulsion and total ionizing dose (TID) are reduced. For example, the plans of the unrealized Asteroid Impact Mission (AIM) included two CubeSats and the Mascot-2 lander \citep{michel2016,herique2016}. 

DISCUS (\cref{fig:discus_6U}) is designed to detect the internal macroporosity of an Itokawa-size \citep{abe_mass_2006} \num{260} to \SI{600}{\m} diameter  {\em Near Earth Asteroid} (NEA) from a few kilometers distance. The primary goal of DISCUS will be to achieve a global signal penetration as well as to analyze the scattering strength for various different penetration depths and measurement positions.  Additionally, forming an actual 3D reconstruction based on the measurements will be attempted. Using the reflected wave, also the shape of the body can be determined. 
Due to the many limitations of space missions, e.g., the strict payload and energy bounds, our radar design aims at a minimal weight and power consumption. Akin to airborne GPR, we apply the stepped-frequency measurement technique and a  half-wavelength  dipole antenna \citep{fu2014}. Maximizing the detectability of deep structures such as voids necessitates using a signal frequency bellow \SI{100}{\MHz}  \citep{francke2009,leucci2008,kofman2012}. We initially target at 20--50 MHz center frequency in the radar design.

This paper is structured as follows. In Section \ref{II}, we motivate the mission plan via a brief review of what is known of rubble pile asteroids. Section \ref{III} describes the on-board scientific instruments and their use. We provide the details of the measurement methods to be applied along with the requirements derived from the known characteristics of the rubble pile asteroids. Section \ref{IV} sketches the mission layout.  An initial design concept of the platform and accessible targets is presented.  Section \ref{V} is devoted to the discussion. Finally, Section \ref{VI} concludes the paper and summarizes the study. 

\section{Rubble pile asteroids in the Solar System}
\label{II}

When the Solar System was formed \SI{4.6}{\giga\years} ago, planetesimals constituted the building blocks of protoplanets and eventually the planets themselves. Comets and asteroids are the remainders of this early stage. Exploring their population is a key to understanding the Solar System's history and evolution, although they contain only about \SI{0.002}{\%} of its total mass.
The asteroids have evolved through several processes; space weathering changed their surface by effects of radiation, internal heating may have altered their  chemistry and mineralogy, and collisional events transformed their interior structure and size distribution. According to the present knowledge, most of the asteroids are actually resulting from catastrophic disruptions of large parent bodies  \citep{farinella_asteroids_1982}. Namely, modeling of catastrophic disruptions \citep{Michel:2001bw,Michel:2002jx} can reproduce the size distribution of the Karin asteroid family, showing that many fragments re-accumulated to form rubble pile asteroids. 

The current evidence also shows that these processes have left most of the asteroids in the size range of about \SI{200}{\m} to \SI{100}{\km} with an interior that is not monolithic but rather an agglomerate of smaller fragments, bound together mainly by gravity and only weak cohesion \citep{richardson_gravitational_2002}. 

\begin{figure*}[t]
        \centering
        \begin{subfigure}[t]{0.48\textwidth}
            \includegraphics[width=\textwidth]{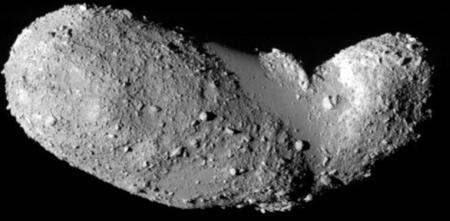}
            \caption{Image of the asteroid Itokawa taken by the Hayabusa spacecraft.
            {Source: JAXA, \url{http://global.jaxa.jp/article/special/hayabusa_sp3/index_e.html} }}
            \label{fig:itokawa1}
        \end{subfigure}
        \vspace{0em}
        \hspace{0.3em}
        \begin{subfigure}[t]{0.48\textwidth}
            \includegraphics[width=1.0\textwidth]{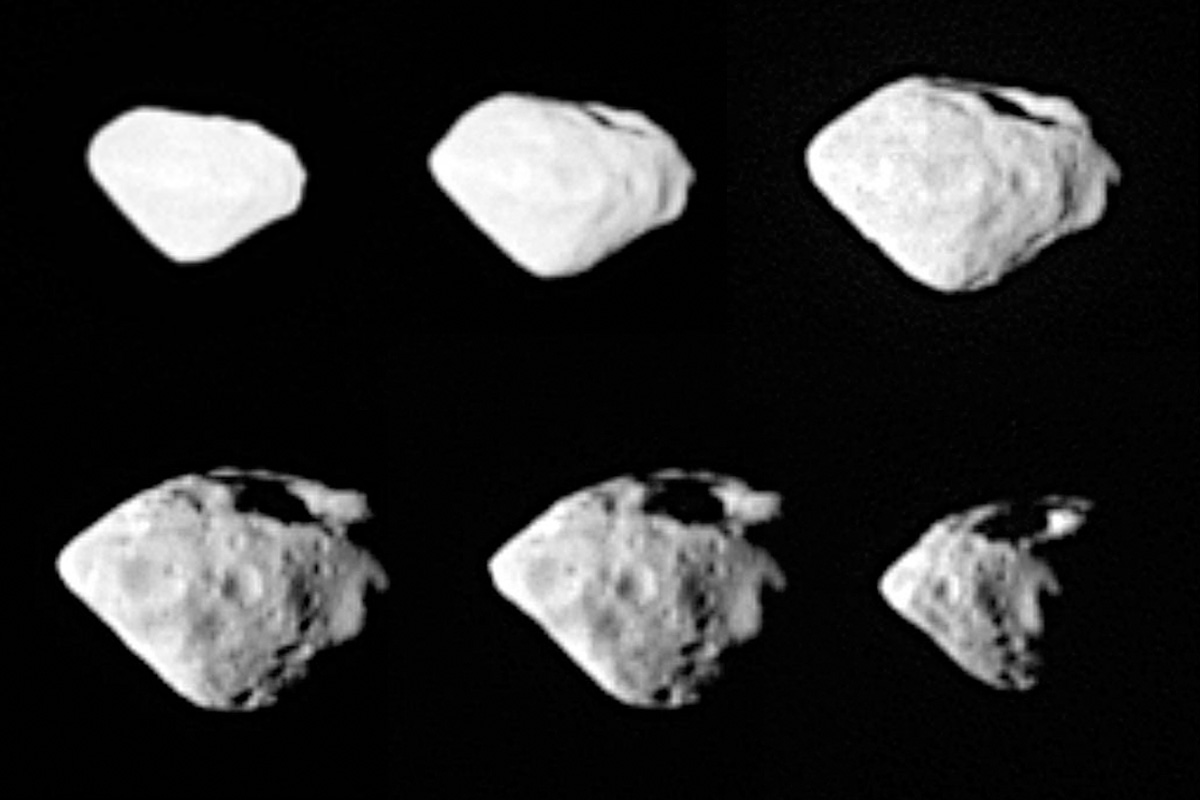}
            \caption{Asteroid \steins{} imaged by the OSIRIS camera system on board of Rosetta.
        {Source: ESA \copyright{} 2008 MPS for OSIRIS Team MPS/UPD/LAM/IAA/RSSD/INTA/ UPM/DASP/IDA}
    }
            \label{fig:steinsdiamond1}
        \end{subfigure}
        \caption[Itokawa and \steins{}, two rubble pile asteroids visited by spacecraft]
        {Itokawa and \steins{}, two rubble pile asteroids visited by spacecraft.}
        \label{fig:rubblepileimages}
    \end{figure*}

The rough comparisons between the meteoritic materials and the indirect density estimates suggest that many asteroids have a large amount of internal macroporosity, that is, voids larger than the typically micrometer sized cracks in the matrix of meteoritic samples (denoted as microporosity). Recently, \cite{carry_density_2012} found high percentages of macroporosity by comparing mass and volume of 287 asteroids, and aggregating information about meteoritic samples linked to their taxonomic classes. For example, the estimate $P_{\text{macro}}=\num{72+-14}$ \% was obtained for asteroid (854)\,Frostia. The measurements provided by the recent Hayabusa and Rosetta missions have also shown that the small asteroid Itokawa \citep[\cref{fig:itokawa1},][]{saito_detailed_2006,abe_mass_2006} as well as the larger one  \steins{} (\cref{fig:steinsdiamond1}) might have around \SI{40}{\%} of macroporosity, respectively. Moreover, simulations of granular flowing material match well with the observed top-like shape of many observed asteroids, e.g., \steins{}; 
simulating the spin-up of rotating asteroids, \citet{Walsh:2008gk} as well as \citet{Sanchez:2012hz} found a material flow towards the equator, forming asteroids with comparable shapes. Analyzing the rotation period as a function of the size has revealed that, for asteroids larger than 200--300 m, there exists a general cut-off at a period of \SI{2.2}{\hours} which only some so-called superfast rotators such as the asteroid 1950 DA\footnote{https://cneos.jpl.nasa.gov/doc/1950da/} surpass  \citep{pravec2002,rozitis_cohesive_2014}. This spin barrier \citep[e.g.][]{pravec_harris_2000,kwiatkowski_photometric_2010} is coinciding with the limit for a cohesionless body before it starts loosing mass at the equator, which indicates that most bodies larger than 200--300 m in diameter are not monolithic in the interior. 

While the evidence for a shattered interior of asteroids is strong, the exact configuration of the interior of these objects is unclear, as direct measurements are still missing. Macroporosity can arise from void space inside the asteroid at various  different scales. There might be large void fractures at the scale of hundreds of meters in the interior, as shown to be likely for \steins \citep[\cref{fig:steinsdiamond1}, see details in][]{deller_hyper-velocity_2017}, or small scale cavities caused by interlocking of boulders. Moreover, the sorting effects of the granular mechanics suggest that the asteroid Itokawa might be a rubble pile with large constituent fragments, explaining the enhancement of the large boulders on its surface. Namely, the amount of regolith observed on Itokawa's surface is significantly larger than what could be produced by impact events forming the observed number of surface craters  \citep{Barnouin-Jha_Itokawa:2008}. Itokawa's  {\em rough highlands} terrain is covered with rubbles following a cumulative power law distribution of \num{-3.1+-0.1}, and the number of large boulders is enhanced compared to models of re-aggregation after a catastrophic impact event, which most likely formed the asteroid \citep{Michikami:2008vg}.  
Indirect inferences of the macroporosity structures can also be obtained through the simulation of the impact events which allows one to connect the crater structure to the surface features and also to determine the shock wave propagation characteristics inside the asteroid. \citet{deller_hyper-velocity_2017} demonstrated this method on asteroid \steins, showing that the catena of pits running from the rim of the crater on \steins{} can be interpreted as a sequence of sinkholes into an interior crack, which has opened due to the impact. 

The mission proposed here provides the means to finally sample the interior of a rubble pile asteroid directly and to provide direct measurements about the distribution of porosity inside these objects. The knowledge obtained would be vital, for example, for the assessment of the asteroid impact deflection as a possible way to neutralize threats by potentially hazardous earth colliding asteroids. Namely, the internal structure is of crucial importance in this regard, since the scale of the porosity inside asteroids determines the propagation of shock waves and, therefore, the efficiency of energy transfer during impact events. Another important application would, obviously, be the mineralogy of asteroids, since the proposed radar measurement technique would provide first-hand information of the dielectric properties in the interior \citep{michel2015}. 

NEAs that are easy to reach for space missions have been explored, e.g., in the Mission Accessible Near-Earth Objects Survey \citep[MANOS,][]{thirouin_mission_2016} focusing on NEAs that have orbital parameters with low \deltav{}, i.e., low relative velocity with respect to the Earth. For a small, but well-described sample of 86 sub-kilometer sized asteroids, it rated seven of these objects as possible targets for manned missions. Using expectations on the number of NEAs from de-biasing studies of the NEA population \citep{tricarico_near-earth_2017}, \citet{thirouin_mission_2016} estimate that a total number of approximately \num{33000} NEAs are accessible for space missions in the \citet{thirouin_mission_2016} definition.

\section{Scientific instrumentation}
\label{III}

\subsection{Stepped-frequency radar}

\begin{figure} 
\centering
  \includegraphics[width=0.48\textwidth]{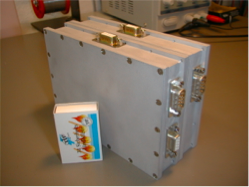} 
\caption{Stepped-frequency radar for the HOPE (Handheld Operated Demining System) \citep{RSTHOPE}. The dimensions of the  box are \SI{12 x 15 x 5}{\cm} and its volume is \SI{900}{\cm\cubed}.}\label{fig:hope_radar}
\end{figure}

The radar design of DISCUS is based on the stepped-frequency technique which  requires a relatively low DC power and low data rate. This is possible since, instead of a single full-bandwidth signal pulse, narrow frequency bands are transmitted and received separately. As the stepped-frequency technique allows measurement of weaker bands, it is generally used in radars with a narrow instantaneous bandwidth as compared to the total bandwidth of the resolution aimed  \citep{lacomme2001}.

\subsubsection{Mathematical concept}

The set of transmitted  narrow frequency lines  $\psi_1, \psi_2, \ldots, \psi_N$ allows one to approximate a given  function $f$ via the sum $f = \sum_{\ell = 1}^{N} c_\ell \psi_\ell$. If $\varphi_\ell$ is the received signal (data) corresponding to frequency line $\psi_\ell$, then the data $g$ resulting from a transmission $f$ is approximately given by $g = \sum_{\ell = 1}^{N} c_\ell \varphi_\ell$, where the coefficients $c_1, c_2, \ldots, c_N$ follow from the formula for $f$. Hence, the data $g$ for any transmission $f$ within the given frequency range can be approximated, if the function pairs $\psi_\ell$ and $\varphi_\ell$ for $\ell = 1, 2, \ldots, N$ are given, i.e., if the stepped-frequency measurement data are available. For a mathematical study of the DISCUS radar, please refer to \citet{takala2017}.

\subsubsection{Baseline design}

The stepped-frequency radar design referred to in this study  was originally developed and verified by {\em RST Radar Systemtechnik AG} within the {European Space Agency's} (ESA's) GINGER ({Guidance and Into-the-Ground Exploration Radar}) project \citep{lentz2000,putz1997}, where a planetary rover was used as the reference platform. In a subsequent ESA project PIRA ({Planetary Into-the-Ground Radar and Altimeter}) this technology was verified under platform movement using a helicopter as the carrier \citep{braun1997}. In addition to space systems, RST has also previous experience on packing a stepped-frequency radar into a very small volume. For example, in the HOPE (Handheld Operated Demining System) landmine detector the total radar volume was \SI{900}{\cm\cubed} (\cref{fig:hope_radar}).

\subsubsection{DISCUS radar}

In the 6U DISCUS design, the goal is to fit a radar with a \SI{2}{\MHz} total signal bandwidth and 20--50 MHz center frequency into a 1U cube (\SI{1000}{\cm\cubed}). The lower end of the targeted frequency  interval is advantageous in order to achieve an appropriate signal penetration \citep{kofman2012} and to minimize solar noise. Namely, if the Sun is active, the dominating error source will be the solar radiation (see \cref{fig:path_loss}) which is on a comparably low level at the targeted interval decreasing towards the lower frequencies  \citep{kraus1966}. The minimum total effect of the active Sun and the galactic background  corresponds to about $20 \pm 5$ MHz, since the galactic noise increases towards the low frequencies. In order to minimize the effect of the Sun, the antenna will be pointed towards it during the measurement. The targeted radar specifications have been listed in  \cref{tab:Radar_Parameters}. 

The duration of a single frequency line can be estimated via $\tau = 1/ B_\ell$, where $B_\ell = B / N$, i.e.,  the total bandwidth $B$ is divided by the number  $N$  of the frequency lines, roughly around 64 lines with pulse duration \SI{32}{\micro\s} can be transmitted and received separately from a 5 km distance to the asteroid (the pulse duration is \SI{96}{\%} of the total  travel-time). A sufficient pulse repetition time (PRT) with respect to the corresponding decay time will be \SI{150}{\micro\s}, i.e., the duration of the complete pulse sequence will be about \SI{9.6}{\milli\s} at the \SI{5}{\km} distance. The narrower is $B_\ell$ the longer is the single pulse duration and the more energy is transmitted. Even when assuming at worst a ground speed of  a few meters per second, a sufficient number frequency lines could be measured. Lower speed would hereby enable to transmit and receive  more lines and enhance the signal-to-noise ratio. The stepped-frequency sequences for larger measurement ranges can also be longer in time and, thereby, contain more lines in order to extend the ambiguity and dynamic range. For each frequency band, the spectrum of the signal is recorded for a time interval determined by the observation distance. 

\begin{table*}
  \centering
    \caption{Targeted radar parameters.}
  \begin{tabular}{ll}
   \toprule
\tabitem Center frequency: &	 20--50 MHz  \\
\tabitem Antenna length: &	Two times 1.5--3.75 m ($\lambda/2$)  \\
\tabitem Radar modulation: & \numrange{32}{2048} lines \\
\tabitem Pulse duration (5 km distance): & \SI{32}{\micro\s} \\
\tabitem Pulse repetition time (5 km distance): & \SI{150}{\micro\s} \\
\tabitem Input/ Transmitting power: & \SI{40}{\W}/ \SI{10}{\W}\\ 
\tabitem Receiver bandwidth: & \SI{2}{\mega\Hz} \\
\tabitem Size:               & \SI{1}{U}\\
\tabitem Weight:			&	\SI{1}{\kg} \\
    \bottomrule
  \end{tabular}
    \label{tab:Radar_Parameters}
\end{table*}

\subsubsection{Link budget and measurement noise}

The initial link budget (\cref{tab:link_budget}) for the received signal is here approximated in decibels via the equation $P_{RX} = P_{TX} + G_{TX} + S + G_{RX}$ where $P_{TX} = 0$ dB (10 W) and $P_{RX}$  denote the transmitted and received power, $S$ the signal power at the receiver location based on an isotropic radiator model,  with an isotropic source and effective antenna aperture $A_{eff} = \lambda^2/(4 \pi)$, and $G_{TX}$  and $G_{RX}$ the gain of the transmitting and receiving half-wavelength dipole antenna, that is, ($G_{TX} = G_{RX} =$ \SI{2.15} {dBi}). 

\begin{figure*} 
\begin{center}
  \begin{minipage}{3.5cm}
\begin{center}
  \includegraphics[width=2.7cm]{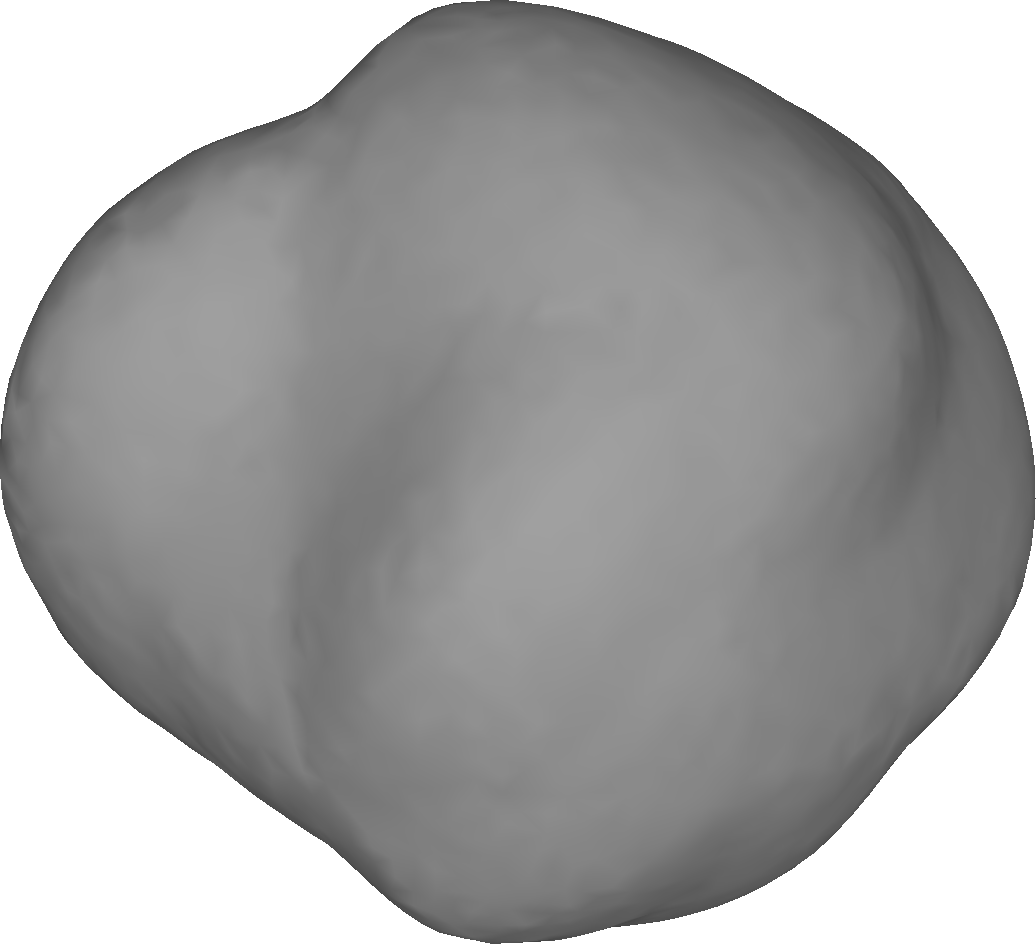}
  \end{center}
  \end{minipage} 
  \begin{minipage}{6.3cm}
\begin{center}
  \includegraphics[width=5.5cm]{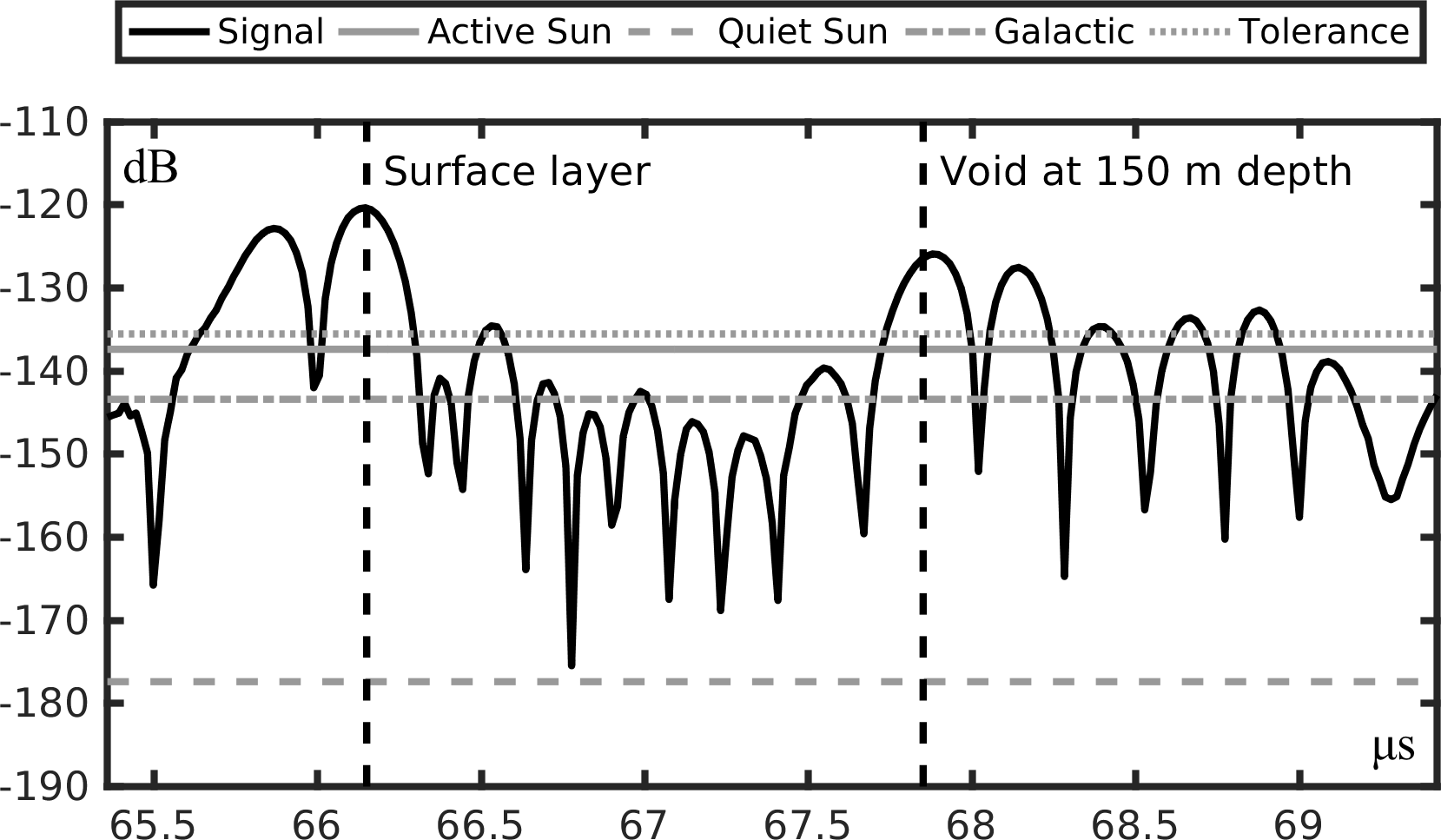} 
  \end{center}
  \end{minipage} 
  \end{center}
\caption{{\bf Left:} The surface model of the asteroid 1998 KY26  scaled to a diameter of \SI{550}{\m} to examine the relative received power $P_{RX}$ for a \SI{10}{\W} signal pulse  with  \SI{20}{\MHz} center frequency and \SI{2}{\MHz} bandwidth \citep{takala2017}. {\bf Right:} The backscattered signal power $P_{RX}$ (dB) received by the transmitting spacecraft at \SI{5}{\km} orbiting distance to the asteroid. The peaks reflecting from a surface layer and a void at 150 m depth have been marked with vertical lines. The noise levels corresponding to active and quiet Sun and the galactic background have been illustrated with horizontal lines. Also, the maximal theoretical noise tolerance -8 dB suggested for a penetrating radar signal \citep{jol2008,erst1984} is shown. The $x$-axis shows the propagation time of the signal ($\mu$s).     \label{fig:path_loss}}
\end{figure*}

In order to obtain an estimate for $S_R$, we simulated the planned radar measurement numerically \citep{takala2017}   using a \SI{550}{\m} diameter synthetic asteroid model based on the surface shape of 1998 KY26\footnote{\url{https://ssd.jpl.nasa.gov/sbdb.cgi?sstr=1998KY26}} (\cref{fig:path_loss}). The center frequency of the measurement was assumed to be  20 MHz, the measurement distance 5 km and the number of frequency lines $N = 64$.  The relative dielectric permittivity distribution  $\varepsilon_r$ enclosed by the surface consisted of a background part ($\varepsilon_r = 4$), a \SI{40}{\m} thick surface layer ($\varepsilon_r = 4$) and three \num{60} to \SI{100}{\m} diameter voids ($\varepsilon_r = 1$) located in the deep interior part. The average signal attenuation rate was around \SI{25}{dB/km} corresponding to a porous basalt at the chosen frequency \citep{kofman2012}. The peak power $P_{RX}$ received for the surface layer and a void in the depth of 150 m was \SI{-120}{\dB} and \SI{-128}{\dB}, respectively. 

The relative measurement noise levels (\cref{tab:link_budget}) caused by the spectral radiation flux density $F$ \citep{kraus1966} from the active and quiet Sun (2E-19 and 2E-23 W/$\hbox{m}^2$/Hz) as well as from the galactic background (5E-20 W/$\hbox{m}^2$/Hz) were estimated via the formula $10 \log_{10} (F B_\ell A_{eff} / P_{TX})$ in which $A_{eff} = 0.13 \lambda^2$ is the effective aperture of the half-wavelength dipole antenna, $B_\ell$ the bandwidth  of a single frequency line, and $P_{TX}$ is expressed in Watts, respectively. Our experience is that the data can be inverted when the total measurement noise level is at least 5--10 dB below the peak level of $P_{RX}$  \citep{takala2017,takala2017c}. In \citep{jol2008,erst1984}, -8 dB has been suggested as the maximal tolerable noise level for penetrating radar measurement. That is, losses  related to penetration, hardware, and signal processing, and miscellaneous reasons can at 5 km distance amount to a total of roughly around 7--15 dB using the galactic background noise as a reference. Due to the two-way signal path and the linear distance dependence of the (stepped-frequency)  pulse length, approaching the asteroid from the distance $d_0$ to $d_1$ would increase $P_{RX}$ in decibels approximately according to the formula $\Delta_{P_{RX}} \approx 30 \log_{10}(d_0/d_1)$ which gives $\Delta_{P_{RX}} \approx +5$ dB for $d_1 = 3.5$ km and $\Delta_{P_{RX}} \approx +9$ dB for $d_1 = 2.5$ km.

\begin{table*}
\centering
   \caption{Initial data link budget  and noise level estimates for 5 km orbiting distance, \SI{20}{\MHz} center frequency, \SI{2}{\MHz} bandwidth,  and \SI{10}{\W} transmission power.}
  \begin{tabular}{lr}
   \toprule
\tabitem Transmitted power $P_{TX}$ & \SI{0}{\dB} \\
\tabitem Half-wavelength dipole antenna gain $G_{TX} = G_{RX}$ & \SI{2.15}{dBi} \\
\tabitem Peak power $P_{RX}$ received    for the surface layer  & \SI{-120}{\dB} \\
\tabitem Peak power $P_{RX}$ received   for a void at 150 m depth  & \SI{-128}{\dB} \\
\tabitem Measurement noise due to  active Sun  & \SI{-137}{\dB} \\
\tabitem Measurement noise due to quiet Sun & \SI{-177}{\dB} \\
\tabitem Measurement noise due to the galactic background   & \SI{-143}{dB} \\
    \bottomrule
  \end{tabular}
    \label{tab:link_budget}
\end{table*}

\subsubsection{Computed radar tomography}\label{Tomography}

Tomographic analysis of the targeted NEA can be performed akin to ordinary tomographic GPR surveys in which radio frequency signals enable reconstructing subsurface ground layers.  Based on the radar specifications, the range resolution ($\delta_r = c/(2 B)$ with $c$ denoting  the signal velocity and $B$ the bandwidth) determining the minimum distance between two separable details  is  \num{20} to \SI{40}{\m} inside the NEA, assuming that its relative electric permittivity is \numrange{3}{12} (e.g.\ solid and pulverized rocks and silica). On ground, the numerical simulations necessary for inverting the radar measurements can be performed using a state-of-the-art computer or computing cluster with sufficient memory for running iterative wave equation solvers within a three-dimensional computation domain. Our recent simulation study demonstrates that a \SI{2}{\MHz} full-wave (full-bandwidth) data can be computed and inverted for the targeted asteroid size range using a high-end workstation computer equipped with up-to-date graphics processing units (GPUs) \citep{takala2017}.  

To record the tomography data, we will use a simple bistatic (two-spacecraft) measurement scenario in which the CubeSats will be placed within a (nearly) constant angle $\gamma$ from each other as seen from the NEA. One of the CubeSats will both transmit and receive the signal and the other one will serve as an additional receiver. As the instrumentation on both CubeSats is identical the concept is redundant by its design. The reference phase can be transmitted by the master satellite. The flying formation does not necessarily need to be maintained stationary: the mutual distance between the CubeSats can slightly vary during the measurements, since their positions will be tracked independently. At least the following two different choices for $\gamma$ will be used: (i) $\gamma \leq \SI{35}{\degree}$ and (ii) $\gamma \geq \SI{145}{\degree}$. The first one of these (i) will enable an enhanced backscattering measurement; our numerical results \citep{takala2017} suggest that with the choice $\gamma = \SI{25}{\degree}$  the accuracy and reliability of the bistatic tomography outcome is superior compared to the monostatic (single-spacecraft) case. The second option (ii) will allow a higher signal-to-noise ratio due to one-way signal paths and tomographic travel-time measurements, performed, e.g., in CONSERT \citep{takala2017b,kofman2015}. 

\subsection{Shape model} 
\label{shape_model}

The altimeter-based shape model of the asteroid can be reconstructed via a combined use of radar and laser  measurements. The laser data will complement the radar measurements in which the first return coming from the surface of the asteroid is detected. Generally, the topographic (flat surface) height accuracy $\delta_{th}$ is below  \SI{20}{\percent} of the range resolution \citep{manasse1960}. Consequently, in empty space, a  measurement with the bandwidth of 2 MHz has the range resolution of $\delta_r = \SI{75}{\m}$ and height accuracy of about $\delta_{th} =  \SI{15}{\m}$. These values are independent of the operation distance. The latter one serves as a rough estimate for the shape model reconstruction accuracy assuming that a sufficiently dense distribution of points over the asteroid surface can be covered at the orbit.

\subsection{Optical imaging}

A  visual camera system is needed to enhance the accuracy of spacecraft positioning.  The images provided by the camera can also be used to optimize the shape model. Because of the limited data transmission rate and the resulting  constrains on the data volume, the angular resolution of the camera has to be chosen to provide a surface coverage with a reasonable amount of images in the expected orbit. Without detailed analyses available at this point, we build our estimate on the experience from the Hayabusa mission \citep{AMICA2010}.
The AMICA imaging system \citep{nakamura2001} on the Hayabusa spacecraft took a total of $>\num{1400}$  images of asteroid Itokawa. The camera had a field-of-view (FOV) of \SI{5.83 x 5.69}{\degree}, and a detector with \SI{1024x1000}{\pixel}. The analog-to-digital converter sampled the signal with 12 bit resolution. This corresponds to an image file size of \SI{1.46}{MiB} uncompressed. Lossless compression can achieve an estimated factor 2 in file size reduction. Therefore, we expect to need to downlink a total of $\approx\SI{1025}{MiB}$ over the course of the mission in order  to achieve a shape model for a scenario similar to Hayabusa. 
Commercial of the shelf camera systems mounted on small satellite platforms achieve a FOV of about \SI{15}{\degree}. While the lower resolution directly decreases the resolution achievable for the surface shape model, the estimate of the data volume is still applicable.  

\section{Mission layout}
\label{IV}

\subsection{Overall design }
\label{od}

For the 6U design of DISCUS, mostly components of the shelf (COTS) have been selected. The spacecraft is illustrated in \cref{fig:discus_6U}. The frame is based on the 6U Supernova Bus of Pumpkin Inc. Electric propulsion will be used. An COTS engine that can provided the required \deltav{} within the power and mass requirements is the Busek BIT-3\footnote{\url{http://www.busek.com/technologies__ion.htm}}. The engine is also used for other CubeSat missions such as the Lunar IceCube \citep{folta2016}.  With a wet weight of \SI{3}{\kg} including \SI{1.5}{\kg} of  propellant, the BIT-3 delivers a \deltav of \SI{3.2}{\km\per\s} for  a \SI{13}{\kg} CubeSat, when operated at \SI{75}{\W} power. 

A Vacco cold-gas thruster and four  star trackers complement the COTS attitude determination and control system. For power generation, the solar panel system MMA eHAWK with \SI{96}{\W} beginning-of-Life (BOL) has been chosen. The camera and a military rated JenOptic altimeter are mounted in the front next to the dipole antenna. Additionally 1U is occupied by housekeeping systems such as battery systems and an on-board computer.
The green and yellow boxes represent the radar and communication system, respectively. The estimated weight of these systems leads to the total wet mass of the satellite, with \SI{20}{\%} margin of around \SI{13.2}{\kg}. The mass budget can be found in \cref{tab:Mass_Budget}.

Since the bistatic measurement process requires accurate timing, e.g., regarding radar pulse window synchronization, a chip-scale atomic clock (CSAC) will be needed on board in order to minimize all possible clocking differences between the two CubeSats \citep{denatale2008}. The synchronization accuracy provided natively by the CSAC will be around 100 ns\footnote{\url{https://www.microsemi.com/document-portal/doc_view/133866-low-noise-csac-datasheet}}, that is, the travel-time corresponding to the maximal attainable radar accuracy  (20 \% of the range resolution, \cref{shape_model}). The precision of the CSAC will allow a several hours' measurement time without re-synchronization, i.e., control pulse exchange.  

\begin{table*} \begin{footnotesize}
  \centering
  \begin{tabular}{lllll}
Component & Specification & Availability & Mass (kg) & Power  (W)\\   
   \toprule
Bus					& 6U w.\ shielding	& modif.\ COTS		& \num{2.0} 	&   \\
Solar panels		& w.\ pointing	  	& announced 		& \num{1.0} 	& \num{96.0}  \\
Propulsion 		 	& Bit 3				& COTS 				& \num{3.0}	& \numrange{56.0}{80} \\
Thruster     		& Cold gas, \SI{50}{Ns}		& COTS 				& \num{0.6} 	& \numrange{0.0}{10.0} \\
Attitude control	& \SI{2}{mNm}		& COTS  			& \num{1.0} 	& \numrange{0.5}{2.0} \\
Housekeeping 		& Incl.\ thermal	& modif.\ COTS 		& \num{1.0}  & \numrange{2.0}{7.0} \\
Radar   			&   				& TRL 3  			& \num{1.0}  & \num{40.0}\\
Dipole antenna  	& $2 \times 3.75$ m		& TRL 2    			& \num{0.4} 	&  \\
Atomic clock		& Chip-Sized Atomic Clock & COTS		&\num{3.5e-2}&\num{0.12}  \\
Nav.\ camera		&      				& modif. COTS 		& \num{0.5}  & \numrange{0.5}{2.0} \\
Laser altimeter		&  					& modif. COTS 		& \num{0.05} & \num{0.01} \\
Communication 		&  					& TRL 2 			& \num{1.5}  & \num{10.0} \\
    \bottomrule
 Total  & & &  \SI{12.1}{\kg} &  \\
 + 20\% Margin & w/o unmodif. COTS & &\SI{13.2}{\kg} & \\ 
  \end{tabular}
   \caption{Mass and power budget for DISCUS. Note that the radar and electric propulsion are not running simultaneously. Power refers to nominal (non-peak) power. The radar power refers to the short stepped-frequency pulse sequences. }
    \label{tab:Mass_Budget}
    \end{footnotesize}
\end{table*}

\subsection{Power requirement}

In the science phase, the total peak power consumption will be about \SI{50}{\W}. The stepped frequency radar will require \SI{40}{\W} non-continuously in short pulses.  The attitude control and housekeeping need together up to \SI{10}{\W}. During the cruise phase, the electric propulsion system requires most of the power. The desired Busek BIT-3 engine can be operated between 55 and \SI{80}{\W}.  The efficiency of the engine will decrease along with the power consumption.

The MMA eHAWK is supplying solar arrays in the required scale. The HaWK 112-7058 delivers \SI{96}{\W} BOL peak power\footnote{\url{http://mmadesignllc.com/existing-hawk-configurations/}}. The system consists, hereby, of two triple folded $2\text{U}\times3\text{U}$ solar panels each mounted on one gimbaled root hinge.  The power consumption estimates have been budgeted in \cref{tab:Mass_Budget}. 

\subsection{Antenna}

Commercial systems using a motor have already proven the feasibility of a \SI{3}{\m} deployable antenna for Cubesats. In example Oxford Space ASTROTUBE BOOM has been validated in space, by unfolding to  \SI{1.5}{\m} \citep{AstroBoom}. The Oxford  system is available for length of up to  \SI{3}{\m} and fits into 1U at a weight of \SI{0.6}{\kg}.   
Our  half-length dipole antenna is projected to be build based on a metal strip design, which is widely utilized for antennas in space applications. A self-enrolling strip has been chosen as the design basis instead of a motor in order to reduce mass and volume of the system. The structure is slightly bent to be more stable and to function as a pre-stressed spring when rolled up. A thermal knife release the tapes, which unwinds entirely by rolling of the spacecraft. First unfolding tests have been successful. Currently, the stability during orbit maneuvers is under investigation.

\subsection{Targets}
\label{targets}

\begin{figure*} 
\centering
  \includegraphics[width=0.90\textwidth]{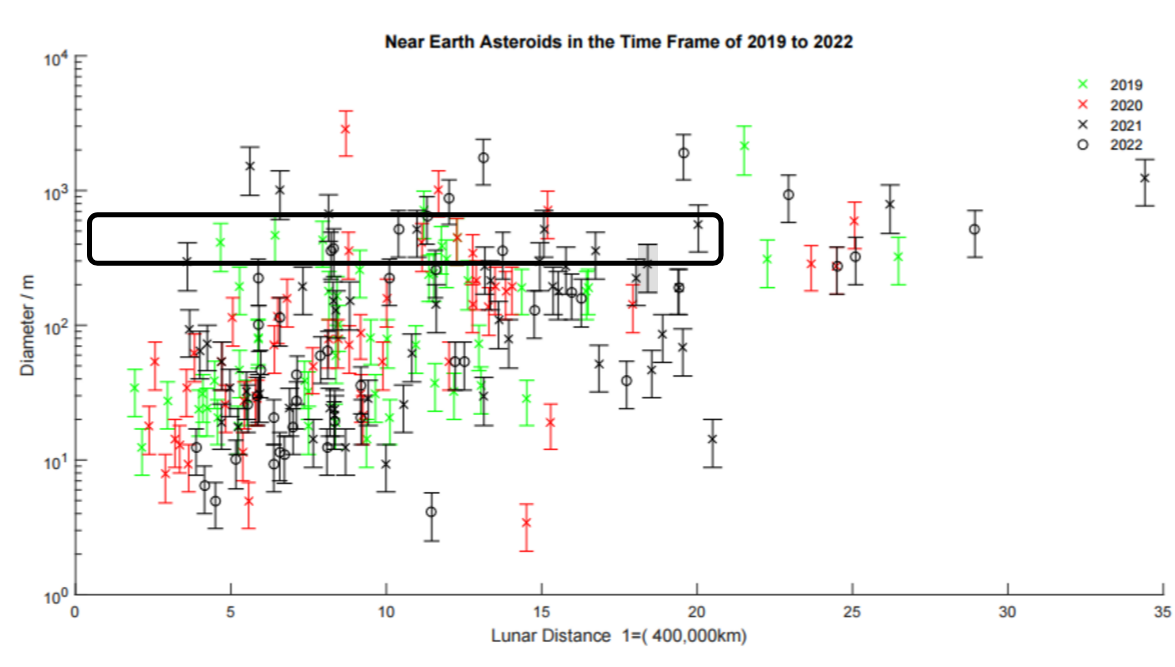}
\caption{Potential targets in the 2010--2022 time frame. The black outlined box shows the most interesting target range with respect to the diameter and the distance of the closest approach. The gray box shows the asteroid 65717 (1993\,B3) which is currently regarded as the most promising candidate with respect to the diameter (\num{180} to \SI{410}{\m}), closest approach distance (18.4~lunar distances) and also \deltav (\SI{3.6}{\km\per\s}).}\label{fig:Ziele}
\end{figure*}

In addition to the scientific and instrument requirements, the target needs to be reachable for a CubeSat. Based on a lunar transfer orbit, a study for an asteroid encountering mission with a 6U CubeSat has been performed by  \cite{JPL_Transfer}. Landis argues that at least an initial \deltav{} of \SI{1.6}{\km\per\s}  into various asteroid orbits could be provided from the lunar transit insertion. Thus, the maximum possible total \deltav will be \SI{4.8}{\km\per\s}, i.e., the sum of the \deltav given by the thruster (\SI{3.2}{\km\per\s}) and that of the lunar transit. As the detailed orbit analysis is still missing, we assume a margin of \SI{0.8}{\km\per\s} and limit our survey to asteroids with a \deltav{} of maximally \SI{4}{\km\per\s}. Still, various asteroids fulfill this requirement.  \Cref{fig:Ziele} shows targets\footnote{\url{https://cneos.jpl.nasa.gov/ca/}} on their closest approach that are reachable on various launch dates. In our opinion, a good reference candidate for 2021 is the asteroid 65717 (1993\,BX3)\footnote{\url{https://ssd.jpl.nasa.gov/sbdb.cgi?sstr=65717}} \footnote{\url{http://www.minorplanetcenter.net/db_search/show_object?utf8=✓&object_id=65717}} because of its close approach of  18.4 lunar distances (LD), relative \deltav{} of around \SI{3.6}{\km\per\s} in January 2021 and diameter between 180 and \SI{410}{\m}\footnote{This size range is based on assumed geometric albedo $p_V$ of 0.05$<$$p_V$$<$0.25. There is an observation by Spitzer Space Telescope
which gives, after radiometric analysis using techniques by \citet{trilling2016}, $D$=$91_{-15}^{+31}$ m and $p_V$=$0.72\pm0.33$ (\url{http://nearearthobjects.nau.edu/}).
However, this albedo solution is unusually high as largest plausible albedos of NEOs are usually not much larger than 0.5 \citep{trilling2010}. Therefore, we
consider this result as a lower limit to the size of 65717 (1993 BX$_3$).}. As the current discovery rate in the recent years has been around 1500~NEAs per year,  more  candidates could show up in the future.  The selection of a mission target has to be revisited once when the final launch window has been fixed.

\subsection{Radar measurement and mapping orbit}
\label{rm}

The two CubeSats will approach the targeted asteroid to a stable point between \SI{5}{\km} and  \SI{10}{\km} altitude and perform their first radar measurements from a static position relative to the target. The moving plane is supposed to avoid shadow phases, which would stress the thermal management and decrease the power generation. If the orbit of the spacecraft is polar, i.e., perpendicular to the spin of the asteroid, a single  orbit will be sufficient to record a backscattering dataset for a nearly uniformly distributed set of measurement points enclosing the targeted NEA. We assume that a polar or nearly-polar polygonal orbit can be achieved with an orbiting direction that is perpendicular to the Solar System's ecliptic plane, since most NEAs are known to have a retrograde spin \citep{la2004}, i.e., the spin orientation is nearly normal to the ecliptic. The orbital movement of the spacecraft will be very slow. That is,  the relative velocity between the asteroid's surface and the spacecraft will be maximally a few meters per second and mainly determined by the spin period.  Consequently, Doppler effects can be omitted. The exact duration of the measurements will depend on several parameters such as the asteroid’s spin period and orbiting distance. For example, a few weeks might be a sufficiently long period.  

For positioning, we will apply the data provided by the radar and laser altimeter, star trackers, and the optical camera. The goal is to obtain an orientation accuracy of at least $3$ degrees relative to the target asteroid's surface, which has been suggested to be sufficient for tomography in \citep{takala2017c}. The asteroid will be visible in the 
star trackers providing an accuracy around 0.01\,degrees  \citep{enright2010}, e.g., with BST's iADCS100. 
The distance to the asteroid can be obtained based on the altimeter data. Using the light-curve inversion, a preliminary estimate for the spin and shape model can be determined during the approach with an estimated couple of degrees accuracy for the spin \citep{durech2015}. As suggested by our present analysis, the shape model can be refined up to an angular accuracy around one degree using the radar at the orbit (\cref{shape_model}). A further refinement will be attempted via optical images. 

In order to ensure that a robust tomographic reconstruction can be produced, we use a 1.5--2.0 spatial oversampling rate with respect to the Nyquist criterion (NC) which was suggested for a monochromatic (travel-time) measurement in \citep{pursiainen2015}. This means recording the radar signal for \numrange{5000}{20000} measurement positions, when extrapolating the measurement positions to the close proximity of a targeted \num{260} to \SI{600}{\m} diameter NEA. The resulting  estimated maximal amount of data recorded by a single satellite will be 250--\SI{1000}{\mega B}, respectively. 
Two identical CubeSats will be present in the mission. A perpendicular orbit will be used for both satellites. The measurement is bistatic, that is, one CubeSat both transmits and receives the signal and the other one serves as an additional receiver. In the measurements, the spacecraft will be placed within a constant angular distance from each other as seen from the asteroid \citep{takala2017}. 

\subsection{Data link}
\label{dl}

The essential prerequisite for the downlink is the ability to transfer the whole scientific dataset during the visibility to one ground-station on Earth. The size of the dataset is estimated to be about \num{12} to \SI{13}{Gibibit}, coming from the radar (\SI{500}{MiB}), the camera (\SI{1025}{MiB}) and an altimeter (\SI{50}{MiB}). The size of the antenna is assumed to be \SI{20x30}{\cm}, the available transmit power is limited by the size of the spacecraft's solar array. Our baseline strategy is to use the \SI{35}{\m} ESA Deep-Space Ground Stations capable to receive in the \num{31.8} to \SI{32.3}{\GHz} band. With a \SI{3}{\W} transmitter RF-power we expect that a data rate of about \SI{520}{Kibibit\per\second} can be achieved  using the DVB-S2 QPSK 2/3 operations mode. With this transfer rate a  dataset of the estimated size can be transmitted to the Earth within less than 7 hours.

\section{Discussion}
\label{V}

In the DISCUS concept, a bistatic (two-spacecraft) low-frequency  radar is carried by two CubeSats. The primary goal is  to resolve the interior structure of a \num{260} to \SI{600}{\m} diameter, i.e., Itokawa-size \citep{abe_mass_2006}, rubble pile asteroid. Based on this study, the most promising target candidate is currently 65717 (1993\,BX3), a potentially hazardous object, which will make its next close Earth approach in 2021 at a \deltav of \SI{3.6}{\km\per\s} with the closest  point being at 18.4~lunar distances from the Earth. More candidates will probably be detected in the near future, and the final target asteroid can only be selected when a launch opportunity has been found.

Akin to airborne GPR, the radar design is based on the stepped-frequency measurement technique coupled with a linear dipole antenna \citep{fu2014}. We aim to use a low 20--50 MHz signal frequency in order to maximize the signal penetration depth  \citep{francke2009,leucci2008,kofman2012} and, thereby, to obtain a global data coverage for the interior part. The main mission objective will be achieved, if the internal macroporosity structures can be detected and mapped. We expect that full or close to full penetration will be achieved, since the level of the signal attenuation in the porous minerals of planetary subsurfaces in the planned low frequency range is known to be low \citep{kofman2012}, e.g.,  \num{10} to \SI{30}{\decibel\per\km} for basalt. Also the outcome of CONSERT shows that signal penetration through an small Solar System body of two times the targeted size range can be achieved \citep{kofman2015}. Moreover, the estimated data transfer capacity (\cref{dl}) allows recording the  signal for a sufficiently dense distribution of points within targeted size range (\cref{rm}).

Our recent simulation study \citep{takala2017} suggests that the tomographic reconstruction of the interior details, such as voids, will be feasible. Based on the results, we expect that the interior can be reconstructed, if the total noise including both measurement and modeling errors is at least \num{5} to \SI{10}{\decibel} below the signal amplitude. This was observed to be the case in CONSERT (comet 67P/Churyumov-Gerasimenko), where the secondary signal peaks were mainly 20 dB below the main peak \citep{kofman2015}. Due to the relatively high  electric permittivity of the asteroid materials we expect a somewhat higher error level for an asteroid. The results of \citep{takala2017}, furthermore, suggest that also limited-angle measurements following from a non-perpendicular (non-polar) orbit will allow obtaining a reconstruction. That is, also NEAs with an exceptional (non-retrograde) spin \citep{la2004} are potential targets for tomographic reconstruction. In the numerical simulations, we have assumed a \SI{25}{\decibel\per\km}  signal attenuation level. In practice, the scattering caused by the macroporosity can lead to some additional attenuation.

The targeted \SI{2}{\MHz} total bandwidth for the stepped-frequency measurements would give the resolution of at least \num{20} to \SI{40}{\m} inside the target asteroid \citep{daniels2005}, if its relative permittivity is \numrange{3}{12} (e.g.\ solid and pulverized rocks and silica). This would mean the distinguishability of the details around the size $1/20$--$1/10$ with respect to the asteroid diameter providing an unforeseen reconstruction  accuracy for the deep interior structure. Consequently, the current design should be sufficient for the main goal of detecting major variances in the volumetric relative electric permittivity structure. For comparison, the initial maximal resolution estimates for CONSERT and {\em Deep Interior}  were 16 and \SI{20}{\m}, respectively \citep{kofman2007,asphaug2003}. In CONSERT, the carrier frequency and bandwidth of the signal were 90 and \SI{8}{\mega\Hz}, respectively. 

We consider the  \SI{2}{\mega\Hz} bandwidth advantageous from both computational and measurement viewpoints, as we currently can invert a full set of data for the targeted asteroid size \citep{takala2017} and sufficient measurement accuracy seems achievable. A wider bandwidth would, however, allow distinguishing smaller details. 
Techniques to increase the bandwidth can, in principle, include adding transmission channels and modifying the antenna folding, loading and increasing the width  \citep{fu2014,poggio1971,balanis2012,alexander2002}. Nevertheless, each of these adjustments would increase the complexity, total weight and power consumption of the system and, therefore, those are not suggested for the initial design. 
Our initial hardware design relies on optimizing the weight and size of the spacecraft in order to minimize the mission costs. Therefore, the 6U frame is used as the reference design.  Based on this study, the \SI{75}{\W} Busek BIT-3 thruster will be able to provide the 6U CubeSat a \deltav of \SI{3.2}{\km\per\s} which together with an initial \deltav obtained from a lunar transfer orbit \citep{JPL_Transfer} will result in a total \deltav of approximately \SI{4}{\km\per\s}.  This would be enough to reach, e.g., the asteroid 65717 (1993\,BX3) in 2021. Depending on the eccentricity of the transfer orbit the targeted \SI{75}{\W} cannot be achieved for the entire transfer. Yet, we expect that including the initial \deltav, the required power for the thruster will have enough margins to compensate these phases. In the current design the total \deltav is limited mainly by the power provided by the solar cells, as they degrade over time and produce less power at sun distances beyond \SI{1}{\au}. The number of reachable targets would increase if solar cells with \SI{120}{W} are used, as the potential of the propulsion system could then be used to the full extend. With currently available solar panels this would the mass by \SI{0.8}{kg}.

A larger 12U size can also be considered, even if all components would fit in the 6U bus and other authors have described the envelope as feasible \citep{NeaScout,6U_Interplanetary}. Namely, the margins are small and could be eaten up by any insertion in the radar, communication system or thermal management. Nevertheless, a 12U bus would lead to a higher total weight and, consequently,  require a much larger orbit control system. As the power would not necessarily scale with size, a completely different propulsion system would be needed. One possibility to overcome the possible issues would be a  piggy-bag ride to an asteroid. This would reduce the need of \deltav{} and communication capability and might lead to an extra 3/2 U of scientific payload, for instance, a more advanced camera system.   

The proposed self-rolled half-wavelength dipole antenna design is beneficial with respect to the mass and volume. We emphasize that the present antenna specifications have been designed for Itokawa-size bodies for which a well-penetrating signal frequency is necessary. Nevertheless, the radar and the spacecraft can also operate with higher frequencies and shorter antenna lengths, which might be preferable for smaller bodies or shallower subsurface structures. In any case COTS solutions at slightly higher weight are available. 

We consider that the goal to fit the radar in a 1U (\SI{1000}{\cm\cubed}) space is feasible based on RST's earlier experience of the \SI{900}{\cm\cubed} HOPE radar design \citep{RSTHOPE}. The next development stage aim to build a \SI{20}{\MHz} radar demonstrator to validate the proposed instrumentation in an on-ground test. RST GmbH has previous experience of such an experiment from ESA's project PIRA \citep{braun1997} in which a stepped-frequency radar equipped with a sub 10 MHz orbiter antenna system was successfully tested under motion.  An ongoing work is to transfer the existing radar design to a below \SI{100}{\MHz} frequency band.

Compared to more frequently used microwave CubeSat radars, for example, Raincube{\textquoteright}s 4U and \SI{10}{\W} Ka-band radar \citep{peral2017}, the planned low-frequency band allows a more simple design. Attaining a considerably better synchronization than what is natively provided by the CSAC, e.g., due to a wider bandwidth, might necessitate extra arrangements. For example, in the recent microwave-range interferometry measurements of the TandDEM-X mission, obtaining a picosecond-level synchronization required relative phase referencing via six dedicated synchronization horn antennas in each spacecraft \citep{krieger2007}.

Future work will include construction of a demonstrator prototype to validate the radar. Numerical simulations concerning the stepped-frequency measurement technique, shape modeling, and also orbit calculations around the asteroid will be performed. We will also examine the capability of the planned instrumentation, e.g., regarding flyby measurements, to reconstruct the interior of a comet and also in mapping of the recently discovered Moon caves. Several details will  need to be refined in the final design stage.

\section{Conclusion}
\label{VI}

DISCUS - The Deep Interior Scanning CubeSat mission is designed with the aim to characterize the interior structure of rubble pile asteroids. Computed radar tomography and a bistatic configuration of two identical CubeSats  will be used to reconstruct internal electric permittivity heterogeneities. While deep-space missions for CubeSat sized spacecraft are still a challenge from an engineering point of view, we found that our radar is compatible with the limitations of a CubeSat mission to a near-Earth asteroid. The baseline design of the DISCUS spacecraft provides a \deltav of \SI{3.2}{\km\per\s}.  A launch to a lunar transfer orbit provides an additional \deltav so that asteroids with a \deltav of approximately \SI{4}{\km\per\s} are reachable. The most prominent target candidate among the currently known NEAs is \num{65717} (1993\,BX3). 

The main scientific result that could be accomplished with DISCUS mission is mapping the distribution of porosity within a rubble pile asteroid. This would be on the one hand of great value to formation models of the Solar System, but also of great importance to deflection scenarios for potentially hazardous asteroids imposing an impact threat to earth. Additionally, crucial knowledge of the internal permittivity structures and, thereby, also the mineral content of the asteroids would be obtained. 

\section*{Acknowledgments}
MT and SP were supported by the Academy of Finland Key Project 305055 and AoF Centre of Excellence in Inverse Problems.\\
PB, JD and EV were supported by the Max Planck Institute for Solar System research.
\bibliographystyle{elsarticle-harv}
\bibliography{CubeSat}

\end{document}